\documentclass[twocolumn]{aastex631}

\newcommand{\be}{\begin{equation}}
\newcommand{\ee}{\end{equation}}

\newcommand{\nudot}{\dot{\nu}}

\usepackage{graphics,graphicx}
\usepackage{epsfig,wrapfig, sidecap} 
\usepackage{amssymb,amsmath}
\usepackage{color}

\begin{document}

\title{NICER timing of the X-ray thermal isolated neutron star RX\,J0806.4--4123}

\author[0000-0003-2317-9747]{B. Posselt}
\affiliation{Oxford Astrophysics, University of Oxford,
Denys Wilkinson Building, Keble Road, 
Oxford OX1 3RH, UK
}
\affiliation{Department of Astronomy \& Astrophysics, Pennsylvania State University, 
525 Davey Lab, 
16802 University Park, PA, USA
}

\author[0000-0002-7481-5259]{G. G. Pavlov}
\affiliation{Department of Astronomy \& Astrophysics, Pennsylvania State University,
525 Davey Lab,
16802 University Park, PA, USA
}

\author[0000-0002-6089-6836]{W. C. G. Ho}
\affiliation{Department of Physics and Astronomy, Haverford College, 370 Lancaster Avenue, Haverford, PA 19041, USA
}

\author[0000-0002-0107-5237]{F. Haberl}
\affiliation{Max-Planck-Institut für extraterrestrische Physik, Giessenbachstraße 1, 85748 Garching, Germany
}




\begin{abstract}
The X-ray thermal isolated neutron star (XTINS) RX\,J0806.4--4123 shows interesting multiwavelength properties that seemingly deviate from those of similar neutron stars. An accurate determination of the spin frequency change over time can assist in interpreting RX\,J0806.4--4123's properties in comparison to those of other XTINSs and the wider pulsar population.      
From 2019 to 2023 we carried out a tailored X-ray timing campaign of RX\,J0806.4--4123 with the NICER instrument.
We used statistical properties of the Fourier coefficients and the $Z_K^2$ test for phase-connecting separate observations and finding a timing solution for the entire dataset. We also developed a simple and universal method for estimating the uncertainties of frequency $\nu$ and its derivative $\nudot$ from the empirical dependencies of $Z_K^2$ on trial values of these parameters, with account of all significant harmonics of the frequency.
Applying this method,
we determined a spin-down rate $\nudot = -7.3(1.2)\times 10^{-17}\,{\rm Hz\, s}^{-1}$. 
The resulting spin-down power $\dot{E}=2.6\times 10^{29}$\,erg\,s$^{-1}$ is the lowest among the XTINSs, and it is a factor of 60 lower than the X-ray luminosity of this neutron star. RX\,J0806.4--4123 is also among the pulsars with the lowest measured $\dot{E}$ in general.

\end{abstract}

%
%
\section{Introduction} 
\label{sec:intro}
RX\,J0806.4--4123 (RX\,J0806 in the following) is a member of the so-called {\emph{Magnificent Seven}}, ROSAT-discovered radio-quiet X-ray thermal isolated neutron stars (XTINSs) that show no obvious non-thermal spectral component in their soft (peak below 1\,keV) X-ray spectrum. However, non-thermal X-ray emission at keV energies was reported by \citet{Yoneyama2017} and \citet{DeGrandis2022} for the brightest and closest member, RX\,J1856.5$-$3754.
Recently, possible new members to this group were discovered with eROSITA by \citet{Kurpas2024,Kurpas2023a}.
The XTINSs do not exhibit any detectable X-ray pulsar wind nebulae (PWNe). 
They have long periods ($P=3-17$\,s) and inferred  surface magnetic dipole fields  on the order of $10^{13}$\,G (e.g., \citealt{Haberl2007,Kerkwijk2007,Pires2014,Hambaryan2017,Pires2019,Malacaria2019}). 
The period derivative $\dot{P}$, and thus the spin-down, was measured with good accuracy for 5 of the 7, 
$ \dot{P}\sim(2.8-11)\times 10^{-14}$ s\,s$^{-1}$
($-\nudot = (0.45$--$4.2)\times 10^{-15}$\,Hz\,s$^{-1}$). 
Because they are close to the magnetars in the $P-\dot{P}$ diagram, the XTINSs are a key population to understand the diversity of neutron star (NS) populations and the evolutionary connections between them. A valid NS evolution model should be able to explain the peculiar properties of the XTINSs. These NSs have characteristic ages that are larger than the kinematic ages by up to a factor ten. 
The X-ray luminosities of the XTINSs are too large for conventional passive cooling, suggesting an additional heating source. The XTINSs also have unusually large (up to a factor 10) optical excess fluxes compared to the long-wavelength extension of their X-ray spectra (e.g., \citealt{Kaplan2011}).  
As shown by \citet{Kaminker2006,Pons2007, Ho2012, Vigano2013} and \citet{DeGrandis2021}, the presence of a strong decaying magnetic field significantly affects the thermal surface emission 
and can account for the high temperatures 
inferred from the X-ray spectra. 
Another explanation for the high X-ray luminosities of the XTINSs is accretion from supernova fallback disks \citep{Alpar2001,Chatterjee2000}. This fallback disk model can also explain the other peculiar properties such as the unrealistically large characteristic ages and the large optical excesses \citep{Ertan2017}.

\begin{deluxetable*}{llrrrc}[ht]
\tablecolumns{6}
\tablecaption{NICER observations of RX\,J0806.4--4123} \label{NicerTab}
\tablewidth{0pt}
\tablehead{
\colhead{ObsID} & \colhead{Year} &  \colhead{Exptime}  & \colhead{TSpan} & \colhead{TCounts} & \colhead{NCR} \\
\colhead{ } &  \colhead{ }  & \colhead{s} & \colhead{s} &  \colhead{ } &  \colhead{cps } 
}
\startdata
2552010101 & 2019.1904  &    9725 & 40 195 & 24681 & $2.273 \pm 0.032$\\
2552010102 & 2019.1918  &   18140 & 84 266 & 46326 & $2.273 \pm 0.031$\\
2552010103 & 2019.1946  &   15600 & 78 808 & 40132 & $2.294 \pm 0.031$\\
2552010104 & 2019.1973  &   10550 & 60 778 & 26945 & $2.238 \pm 0.032$\\
mergeA1    & 2019.1904  &   54015 & 278 038 & 138083 \\
\hline
2552010201 & 2019.1995  &    3550 & 12 361 & 9228 & $2.313 \pm 0.040$\\
2552010202 & 2019.2001  &   13050 & 35 413 & 34276 & $2.265 \pm 0.033$\\
mergeA2    & 2019.1995  &  16600  & 51 281 &  43504 & \\
\hline
2552010301 & 2019.2304  &   13330 & 73 602 & 35034 & $2.186 \pm 0.032$\\
2552010302 & 2019.2329  &    9138 & 62 127 & 23976  & $2.176 \pm 0.033$\\
mergeA3    & 2019.2304  &    22468    & 140 093 & 59010 \\
\hline
2552010401 & 2019.4029  &   18510 & 68 771 & 48449 & $2.205 \pm 0.032$\\
A4         & 2019.4029  &   18510 & 68 771 & 48449 & $2.205 \pm 0.032$ \\
\hline
2552010501 & 2020.1472  &     840 &    869 & 2323 &  $2.206 \pm 0.069$\\
2552010502 & 2020.1476  &   16180 & 84 433 & 42467 & $2.186 \pm 0.036$\\
mergeA5    & 2019.1472  &   17020 & 94 575 & 44790 &  \\
\hline
mergeA     & 2019.1904  &  128613 & 30 281 952 & 333837 & $2.312 \pm 0.030$ \\
\hline
3553010101 & 2021.1365  &    3871 & 12 585 & 4684 & $2.405 \pm 0.058$\\
3553010102 & 2021.1370  &   14300 & 80 579 & 39031 & $2.105 \pm 0.038$\\
3553010103 & 2021.1397  &    5933 & 28 889 & 16169 & $2.242 \pm 0.039$\\
\hline
mergeB     & 2021.1365  &   21820 & 128 231 & 59884  &   $2.314 \pm 0.032$\\
\hline
5630010101 & 2023.0690  &   14760 & 66 901 & 37145 & $2.171 \pm 0.031$\\
5630010102 & 2023.0712  &    9197 & 85 095 & 23758 & $2.154 \pm 0.035$\\
5630010103 & 2023.0743  &    363  &  5 969 & 1016 &  $2.134 \pm 0.092$\\
5630010104 & 2023.0928  &    1331 &  6 236 & 3182 &  $2.074 \pm 0.049$\\
5630010105 & 2023.0933  &    4976 & 28 619 & 12313 & $2.159 \pm 0.034$\\
5630010106 & 2023.1011  &    450  &   449  & 1145 &  $2.195 \pm 0.077$\\
5630010107 & 2023.1016  &    792  &  5 820 & 2107 &  $2.269 \pm 0.060$ \\
\hline
mergeC     & 2023.0690  &   31870 & 1 034 976 & 80666  & $2.247 \pm 0.030$\\
\hline
\enddata
\tablecomments{For each observation with identification number (ObsID) the starting time in the NICER archive is listed in decimal year. 
The  third column, Exptime, indicates the effective exposure times (obtained from spectra via XSPEC). 
The fourth  and fifth columns list the covered time span of each (cleaned) timing data set and the number of total (source + background) counts using an energy filter of 0.23--1.0 keV band (PI between 23 and 100).  XSPEC-based net count rates, NCR, 
were obtained with the SKORPEON background model and are reported for an energy filter 0.23--1.0\,keV to illustrate consistent source flux level even if the background varied. 
\label{NicerTab}}
\end{deluxetable*}

At the position of RX\,J0806, \citet{Posselt2018} detected \emph{extended} near-infared (NIR) emission with the \emph{Hubble} Space Telescope. This NIR-emission can be explained by either a disk or an unusual PWN (there are no other known NIR-only PWNe). 
Although NIR limits for the other XTINSs are less deep than for RX\,J0806, there is also the question whether RX\,J0806 may be different, for instance in its timing properties.
The period of RXJ0806 is $P\simeq 11.37$\,s ($\nu\simeq 0.0879$\,Hz) \citep{Haberl2002}. 
Using XMM-Newton observations of 2008--2009, \citet{Kaplan2009} 
reported $\dot{\nu}_{-16}\equiv \dot{\nu}/(10^{-16}\,{\rm Hz\,s}^{-1}) = -4.3 \pm 2.3 (1\sigma)$.
Because the presence of putative fallback disk can be accompanied by
 substantially different timing properties of this INS, a more precise constraint on $\dot{P}$ is highly desirable to assess whether RX\,J0806 is perhaps an exception among the XTINSs.
For this reason, we carried out a multi-year NICER timing campaign of this neutron star.

\section{Observations and Data reduction} 
\label{sec:obs}
For our timing program we used the planning approach outlined in Appendix~\ref{sec:planning}. 
We started our timing program of J0806 in 2019/2020 with five observation epochs 
separated by increasing time spans to optimize time coverage and phase connection (Program ID 2552). We continued with one observing epoch in 2021, and another observing epoch in 2023 (Program IDs 3553, 5630).   
In total, NICER data from 21 observations
were acquired from 2019 to 2023, covering a total of about 3.9 years.
Table~\ref{NicerTab} lists the parameters of the individual NICER observations.\\

For data analysis, we used the NICERDAS software (version 2023-08-22\_V011a)
with HEAsoft 
(version 6.32 \citealt{HEAsoft2014}) 
and the current version of the calibration files (CALDB xti20221001).
After initial processing of individual observations with standard calibration and filtering, lightcurves in different energy ranges (Pulse Invariant, PI, channel ranges $30-150$\footnote{PI $30-150$ corresponds to $0.3-1.5$\,keV, see \href{https://heasarc.gsfc.nasa.gov/docs/nicer/analysis\_threads/gain-cal/}{heasarc.gsfc.nasa.gov/docs/nicer/analysis$\_$threads/gain-cal/}}, $150-800$, $800-1500$)
with different binnings (1\,s, 30\,s, 60\,s) were produced using the task \texttt{nicerl3-lc}. Evaluating these (noisy) lightcurves, we defined two good-time-interval (GTI) requirements as count rate $<0.5$\,counts\,s$^{-1}$ (strict GTI filter) and count rate $<1$\,counts\,s$^{-1}$, based on the background lightcurve in the PI-range $800- 1500$ with time bin 30\,s.   
The (updated) lightcurves were inspected for peculiarities.
Previous versions of the NICERDAS software required additional manual GTI selection, avoiding suspiciously low count rates (at least a factor 3 less than the average), e.g., in ObsID 3553010102 in 2021. With 2023-08-22\_V011a, this was no longer necessary.
We checked the signal to noise level in the timing analysis (Section\,\ref{sec:timing}), and decided to use the strict filter for our observations except for year 2021. As these data from our second year were noisier,
we had to use the less strict GTI filter. For estimates of the effective exposure times and net count rates in Table~\ref{NicerTab}, we produced spectra from the GTI-filtered data with the SKORPEON background model using the task \texttt{nicerl3-spect}.
We used XSPEC (version 12.13.1) to estimate effective exposure times and net count rates.

\begin{deluxetable*}{rccrrrrr}
\tablecolumns{7}
\tablecaption{Estimates of $\nu$ and $\nudot$ based on $Z_{K,\rm max}^2$} 
\tablehead{
\colhead{epoch} & \colhead{MJDmid} & 
\colhead{$Z^2_{1\rm{, max}}$} & \colhead{$Z^2_{2\rm{, max}}$} & \colhead{$Z^2_{3\rm{, max}}$} & \colhead{$\nu - 0.0879477$\,Hz} & \colhead{$\dot{\nu}$} \\
\colhead{} &  \colhead{} &  
\colhead{} &\colhead{} &  \colhead{}  & \colhead{$10^{-8}$\,Hz} & \colhead{$10^{-17}$ Hz/Hz}\\
}
\startdata
mergeA1  & 58555.0   &145 & 160 & 164 & $-9  \pm  24$ & $\cdots$\\
mergeA2  & 58557.2   & 63 &  65 &  67 & $-127 \pm  163$ & $\cdots$\\
mergeA3  & 58568.8   & 66 &  76 &  76 & $3   \pm  59$ & $\cdots$\\
A4       & 58631.4   & 61 &  76 &  78 & $-94  \pm  97$ & $\cdots$\\
mergeA5  & 58903.4   & 65 &  72 &  81 & $-138 \pm  106$ & $\cdots$\\
mergeB  & 59265.6   & 88 &  92 &  96 & $-49  \pm  52$ & $\cdots$\\
mergeC  & 59973.0   & 90 & 104 & 108 & $3    \pm  6$ & $\cdots$\\
\hline
A1+mergeA2   & 58555.5  & 201& 219 & 221 & $3.2 \pm 20  $   & $\cdots$\\
...+mergeA3  & 58558.8  & 266& 294 & 295 & $3.2 \pm 3.6  $  & $\cdots$\\
...+A4       & 58571.0  & 327& 366 & 366 & $3.25 \pm 0.74 $ & $\cdots$\\
...+mergeA5  & 58615.6  & 389& 433 & 436 & $3.69 \pm 0.17 $ & $1 \pm  10$\\ 
...+mergeB  & 58714.4  & 473& 523 & 526  & $3.57 \pm 0.08 $ & $-9.8 \pm 3.1 $\\
...+mergeC  & 58928.4  & 562& 619 & 626  & $3.44 \pm 0.04 $ & $-8.0 \pm 0.9 $\\
\enddata
\tablecomments{The first part of the table shows the $Z_{K,\rm max}^2$ statistics for individual data clusters separately, while in the second part of the table one more cluster is added to each consecutive line (see text for more details).  
The times of the middle of the considered events in an epoch (i.e., the reference time for the individual calculations) are shown as (rounded) MJDmid.
The $\nu$ and $\dot{\nu}$ values correspond to maxima of $Z_2^1$. We list $1\sigma$ uncertainties. \label{Z2ks}}
\end{deluxetable*}

\begin{figure*}[t] 
 \includegraphics[width=9.2cm]{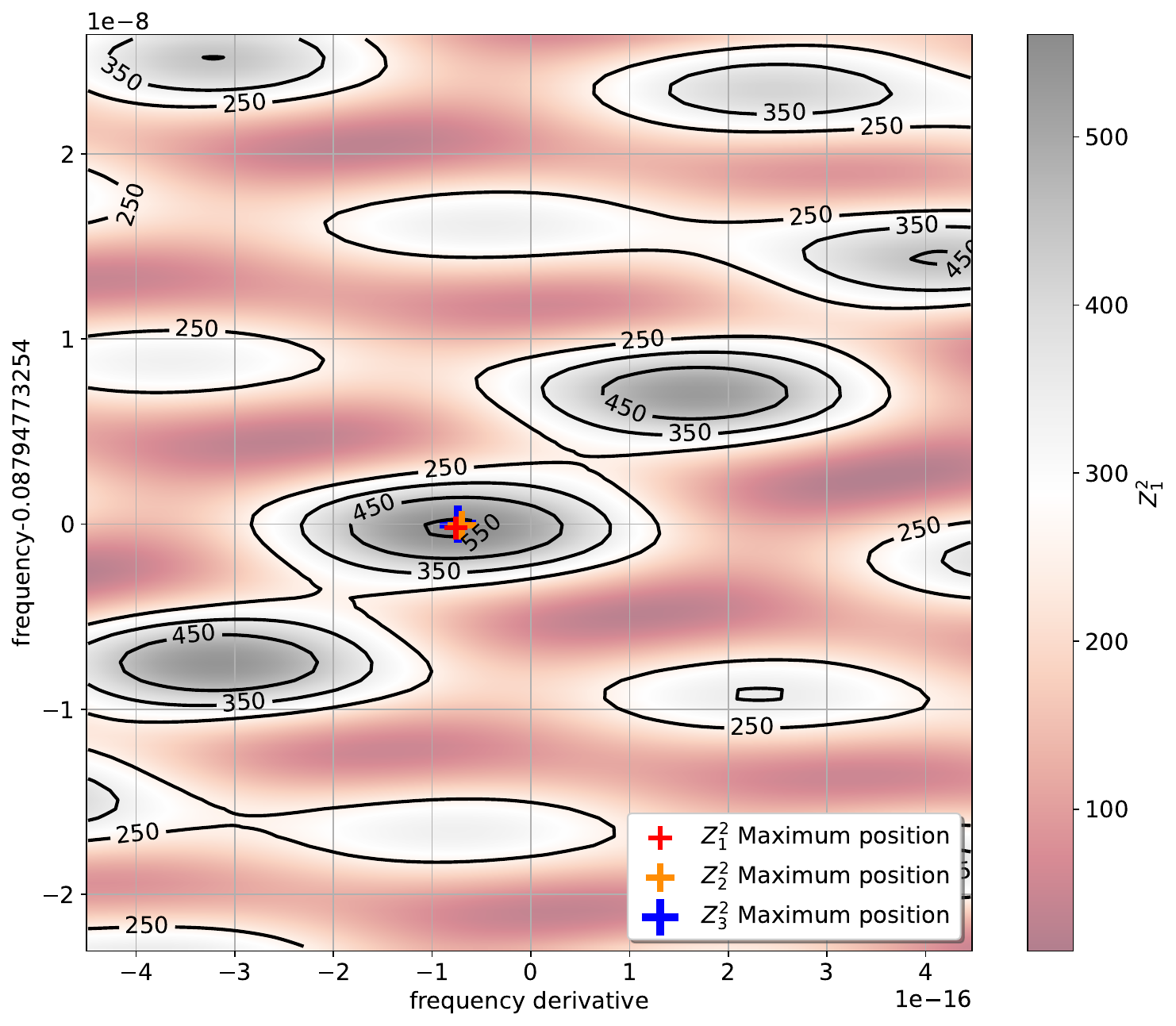}
 \includegraphics[width=9.2cm]{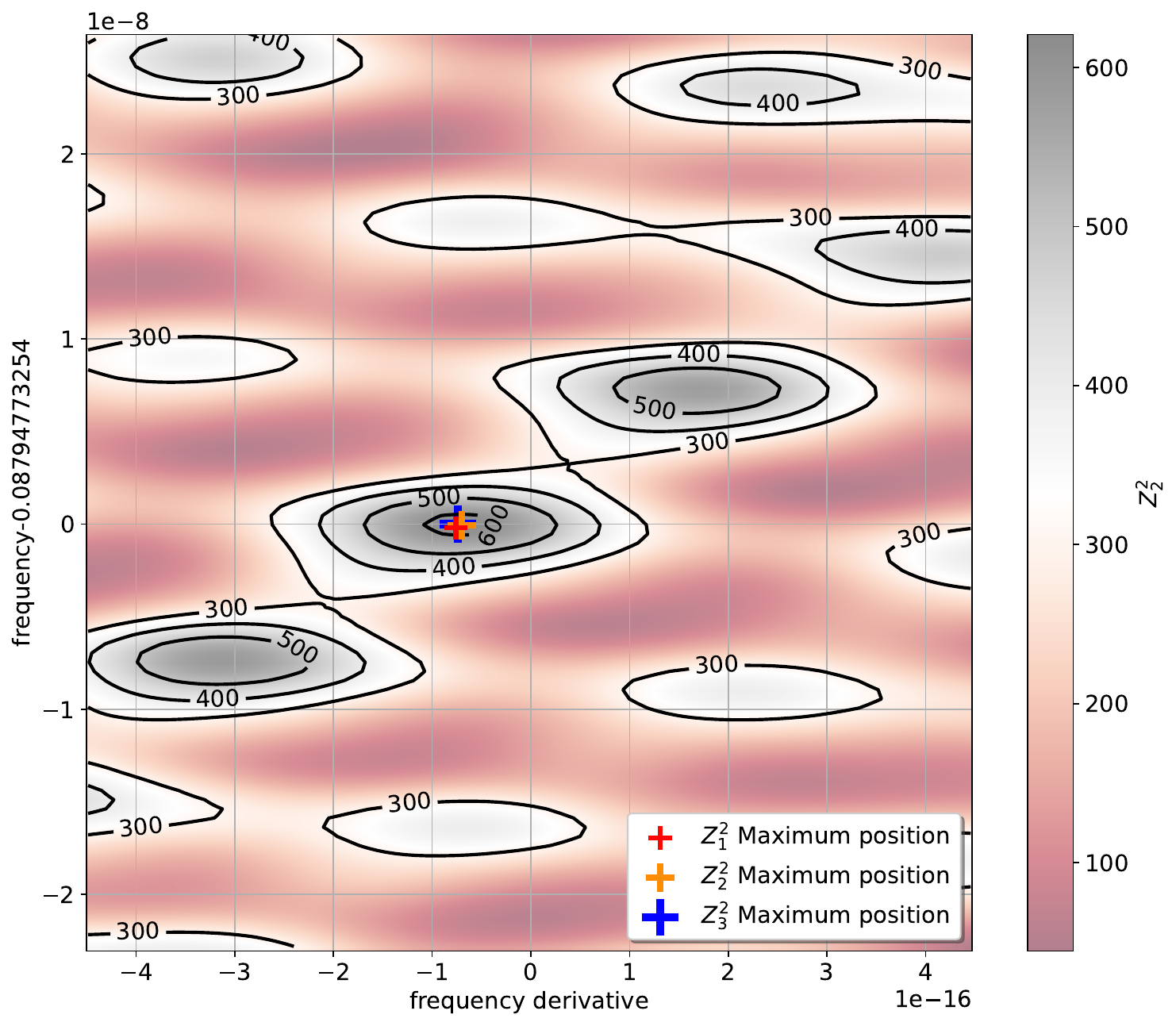}
\caption{The test statistics  $Z^2_{1}$ (left panel) and $Z^2_{2}$ (right panel) in the $\dot{\nu}-\nu$ plane. The maximum values are indicated by crosses. These plots are based on a sampling of $\Delta \nu=5 \times 10^{-10}$\,Hz and $\Delta \dot{\nu}=3 \times 10^{-18}$\,Hz/s, and they are smoothed using bilinear interpolation. 
\label{nunudotplanewide}}
\end{figure*}

\section{Timing analysis} 
\label{sec:timing}
There are different methods to measure timing parameters, such as $\nu$ and $\nudot$, for a pulsar.
The commonly used approach in radio pulsar timing connects the phases of pulse times-of-arrival (ToA).
This involves creating a ``folded'' pulse profile from many individual pulses and 
searching for the best timing parameters
by minimization of differences between phases of pulse maxima in several segments of the observation, or groups of observations (for details, see, e.g., \citealt{Taylor2024, pulsarHB2004,Ransom2002}).
Since, unlike the radio timing, X-ray pulsar timing is based on the detection of separate photons (events), 
a pulse ToA approach requires phase binning, which introduces additional uncertainties. Therefore, it is preferable to use {\em unbinned data}, i.e., {\em event} ToA instead of pulse ToA. In contrast to most radio pulsars, the thermal X-ray pulses from XTINSs, including RX\,J0806, are very broad and smooth, with a small number of contributing harmonics. A convenient approach to the timing analysis of unbinned data of such pulsations involves the phases and amplitudes of Fourier harmonics calculated with the aid of sums over the events (see Appendix A.1 and references therein).
If the pulsations are stable throughout the analyzed time interval, the most probable timing solution 
for several groups of observations corresponds to the best phase connection between these groups, i.e., 
it minimizes
the differences between the harmonic phases of the separate observations. As we show in Appendix B, minimizing the phase differences is equivalent to maximizing the well-known $Z_K^2$ statistic for that group of observations\footnote{This was also demonstrated by \citet{Halpern2011} for a specific example.}, where $Z_K^2$ is the sum of powers of $K$ harmonics that characterize the pulse profile 
-- see Appendix~\ref{subsec:general_expressions}. 

The $Z_K^2$ statistic is calculated on a parameter grid (e.g, a 1D frequency grid or a 2D $\nu$-$\nudot$ grid).
The grid computations can be time-consuming if there is a lack of prior information on these parameters, particularly when probing high frequencies. In our case, however, the pulsation frequency, $\nu\approx 0.0879$ Hz, is small and known with a reasonable precision, and the frequency derivative is constrained, $|\nudot| \lesssim 10^{-15}$ Hz\,s$^{-1}$. Therefore, we chose the $Z_K^2$ approach for our timing analysis of RX\,J0806, with the main goal to confidently measure the frequency derivative.

The 21 NICER observations are listed in Table~\ref{NicerTab}.
There are numerous gaps in the data (Exptime $<$ TSpan in Table~\ref{NicerTab}) as is normal for NICER observations.
Examination showed that some of the individual observations have too few counts to detect the pulsations.
Hence, we merged the data  from 21 observations into 7 ``clusters'' (A1 through A5, B, and C), as indicated in Table~\ref{NicerTab}, and confirmed their phase-connection within each cluster by evaluating the aliases in $\nu$, caused by the gaps (see Section \ref{sec:mult_obs}).
For computational efficiency, we adapt our grid cell sizes
$\Delta \nu$ and $\Delta \dot{\nu}$ (specified in the respective text) to the improving time coverage and analysis goal.

We start our $\nu$, $\nudot$ measurement from a relatively crude analysis, deferring a refined measurement to the next step. 
For each analyzed dataset, we used the middle of all respective event times as the reference time.
We calculated the $Z_K^2$ values on the $\nu$-$\dot{\nu}$ grid with step sizes 
$\Delta \nu=5\times 10^{-10}$\,Hz, and changing step sizes for $\Delta \dot{\nu}$ depending on number of considered epochs 
($\Delta \dot{\nu}$ values range from $2\times 10^{-17}$ Hz s$^{-1}$ to 
$3\times 10^{-18}$ Hz s$^{-1}$). 
With the aid of 
the H-test \citep{Jager1989}, 
we estimated that not more than three Fourier harmonics can give statistically significant contributions.

The results of the crude analysis are provided in Table~\ref{Z2ks}. In the first 7 lines of Table~\ref{Z2ks} we report the values of $Z_K^2$ maxima and the corresponding frequencies for each of the 7 clusters. 
In accordance with our observational plan, the first cluster A1 has
the largest exposure time and correspondingly largest individual $Z_K^2$. Hence, we will use it as the 
reference cluster in the next steps.
Proceeding to the next epochs, we consecutively added each following cluster and 
calculated the $Z_K^2$ maxima, frequencies, and frequency derivative (when measurable) -- see the last 6 lines in Table~\ref{Z2ks}. 
Uncertainties of $\nu$ and $\dot{\nu}$ 
are estimated 
using the ``empirical approach'' described in Equations~(\ref{eq:empirical_uncertainties}) and (\ref{eq:g-factor}).
The addition of each new cluster narrowed the range of allowed frequencies. Starting from the merged dataset A+B, the timing data become sensitive to $\nudot$.
The values of $\nu$ and $\dot{\nu}$ obtained from $Z^2_{2\rm{, max}}$ 
(and $Z^2_{3\rm{, max}}$) agree with the $Z^2_{1\rm{, max}}$-based values within $1\sigma$ of the uncertainties. 

Because of the multiple observational gaps, the $Z_K^2(\nu)$ and $Z_K^2 (\nu, \dot{\nu}$) dependencies contain large numbers of 1D and 2D peaks, respectively.
If one of the peaks is substantially
higher than the others, this ``main'' peak corresponds to the true solution while the other peaks 
correspond to aliases  (see Sec.\ \ref{sec:mult_obs} and Appendix \ref{sec:planning}). 
If the height(s) of the peak(s) in the vicinity of the heighest one were only slightly lower, one 
would not be able to determine the peak corresponding to the correct solution.
The better is phase connection between the separate observations (clusters), the lower are the alias peaks in comparison with the main peak. Since the durations of the separate observations and the gaps between them were chosen from the requirement of good phase connection (see Appendix C), the alias peaks are indeed considerably lower than the main peak.
While adding new clusters, we made sure that we tracked the main peak and avoided any aliases. A 2D example (all clusters added) is shown in Figure~\ref{nunudotplanewide}. 
In our case, the next peak in height with respect to $Z^2_{1\rm{, max}}$ ($Z^2_{2\rm{, max}}$, $Z^2_{3\rm{, max}}$) has a 
$\dot{\nu} \sim -3.1 \times 10^{-16}$ Hz s$^{-1}$ and is significantly lower, by 
$\Delta Z^2_{1}=18$ ($\Delta Z^2_{2}=32$, $\Delta Z^2_{3}=37$).\\ 

The time of the first used event in our merged data set has a time stamp of MJD (TDB\footnote{TDB stands for Barycentric Dynamical Time at the Solar System Barycenter, the time frame in which we carry out the timing analysis.}) 58553.5031368.
and the full time coverage is $\approx 1429$\,days.
Adding $680.0$\,days to this first event time defines the reference time for our timing analysis.
This reference time allowed us to minimize correlations between $\nu$ and $\dot{\nu}$, enabling reasonably independent uncertainty estimates according to Appendix~\ref{sec:nu-nudot_uncertainties}.

\begin{figure}[b] 
\includegraphics[width=9.3cm]{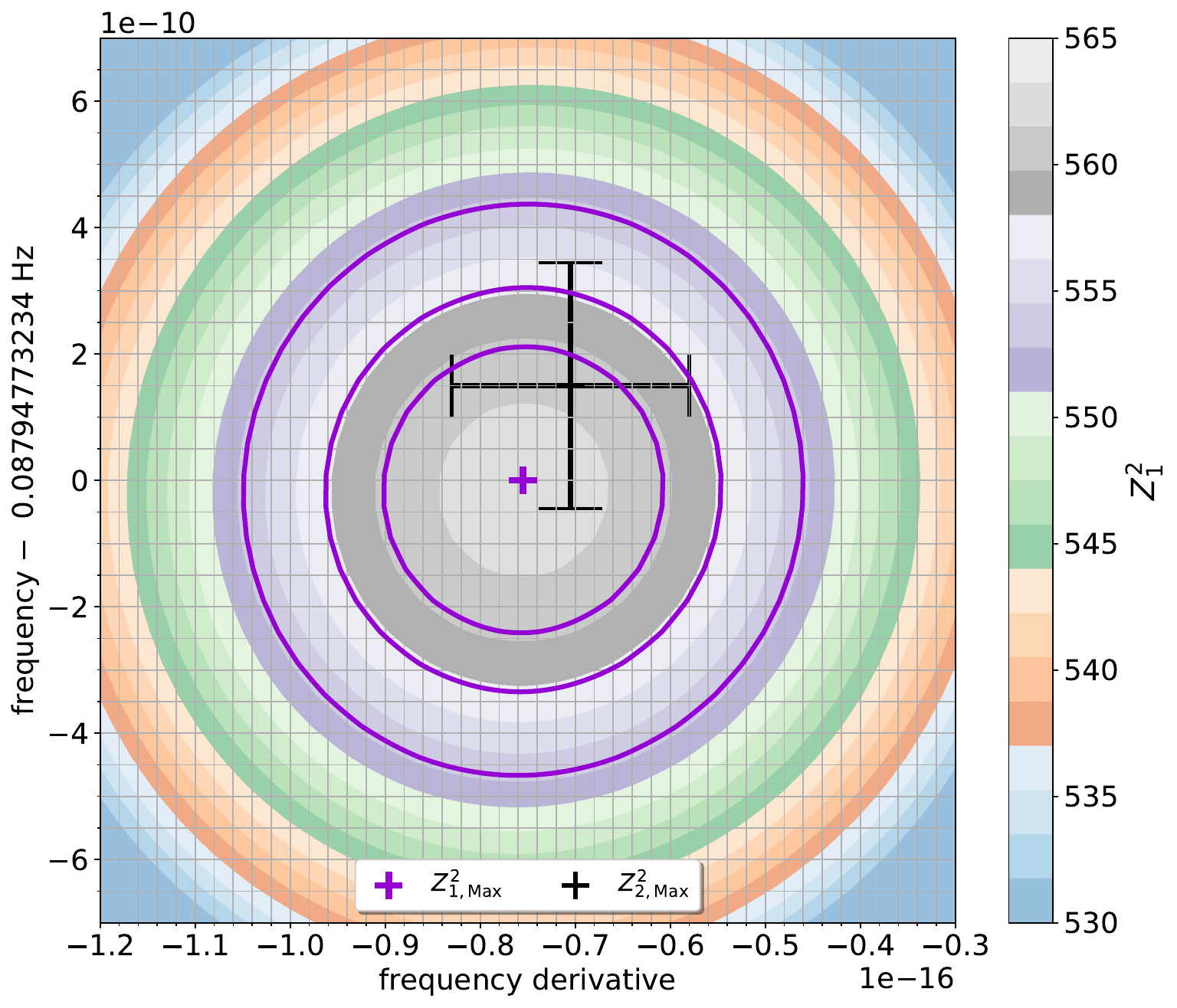}
\caption{The test statistics  $Z^2_{1}$ in the $\nu$-$\dot{\nu}$ plane. Our 4 years NICER data coverage of J0806 are considered with a reference time of 680\,days from the first detected photon. The small cross indicates the respective maximum value, surrounded by its 68\%, 90\% and 99\% confidence contours (calculated as described in Appendix~\ref{sec:nu-nudot_uncertainties}). The larger black cross marks the 68\% confidence regions of $Z^2_{2}$, respectively. \label{nunudotplane}}
\end{figure}

Based on the first estimates in Table~\ref{Z2ks}, we then employ a finer sampling of $\Delta \nu=5\times 10^{-11}$\,Hz and $\Delta \dot{\nu}= 5\times 10^{-19}$ Hz s$^{-1}$ to investigate the highest peaks in more detail, and to determine the 68\%, 90\%, and 99\% confidence contours for two parameters of interest in the $\nu$-$\dot{\nu}$ plane, see Figure~\ref{nunudotplane} for the highest $Z^2_{1}$ peak. 
The values of $Z^2_{1\rm{, max}}=562$, $Z^2_{2\rm{, max}}=621$, $Z^2_{3\rm{, max}}=627$ differ slightly from the values (562, 619, 626, respectively) in Table~\ref{Z2ks} due to the finer sampling and different reference times.
Using Figure~\ref{nunudotplane} for the 68\% confidence level (for two parameters of interest), 
our timing solution is:
\be
\label{timsol}
\begin{split}
\nu=0.087 947 732 54 (18)\,{\rm Hz} \\ 
\nudot = -7.3(1.2)\times 10^{-17}\,{\rm Hz\, s}^{-1}
\end{split}
\ee
at the reference epoch MJD (TDB) 59233.5031368.
\begin{figure}[b!] 
\hspace{-0.5cm}
\includegraphics[width=8.7cm]{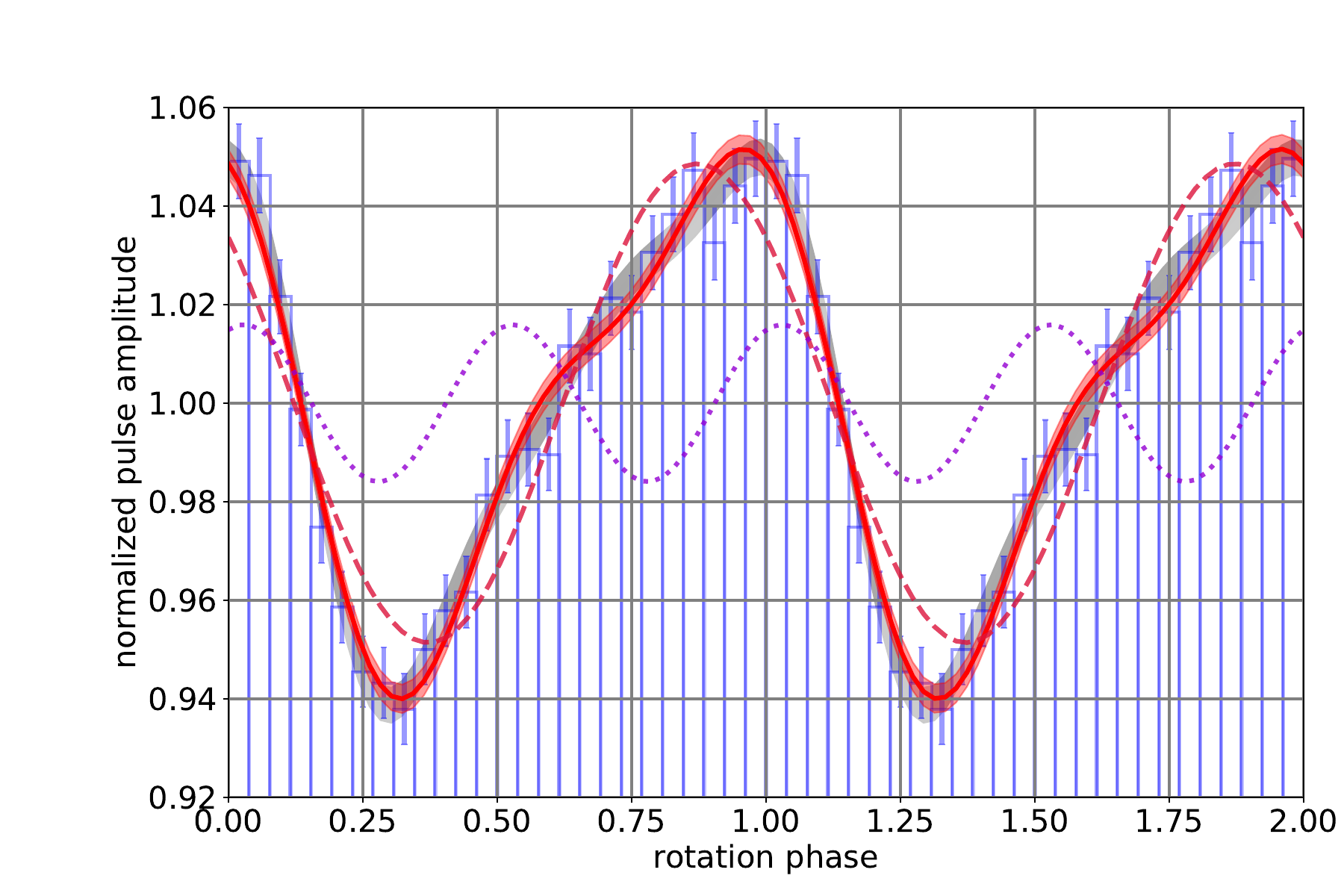}
 \caption{The pulse shape of J0806.  
 The zero phase corresponds to the reference time MJD (TDB) 59233.5031368. 
 The blue histogram shows the folded X-ray counts in the energy range $0.23-1$\,keV.
The red solid line and red shaded region show the pulse shape as the sum of two harmonics and its $1 \sigma$ uncertainty range. The contributions of the first (dashed red) and second (dotted violet) harmonics  are also indicated. For comparison, the shaded grey profile demonstrates how the pulse shape would change if the (insignificant) third harmonic is included. 
 \label{pulseshape}}
\end{figure}

For further analysis of our timing data in the energy range $0.23-1.0$\,keV, we follow the definitions of the Fourier coefficients in  Appendix\,\ref{subsec:general_expressions}.
The harmonic's amplitude $s_K$  and phase $\psi_K$ are listed in Table~\ref{skpsik} for each of the three merged epochs (A, B, C), and for the fully merged data (A+B+C). For the third harmonic,
we find an amplitude $s_3=0.0053 \pm 0.0029$, i.e., 
this harmonic is statistically insignificant.
Comparing the pulse shapes composed of two and three harmonics with the binned data and with each other (see  Figure~\ref{pulseshape}) also illustrates 
the negligible contribution from the third harmonic, which we ignore for the further analysis. 

\begin{figure*}[] 
\vspace{-0.5cm}
\hspace{-0.5cm}
\includegraphics[height=9.1cm]{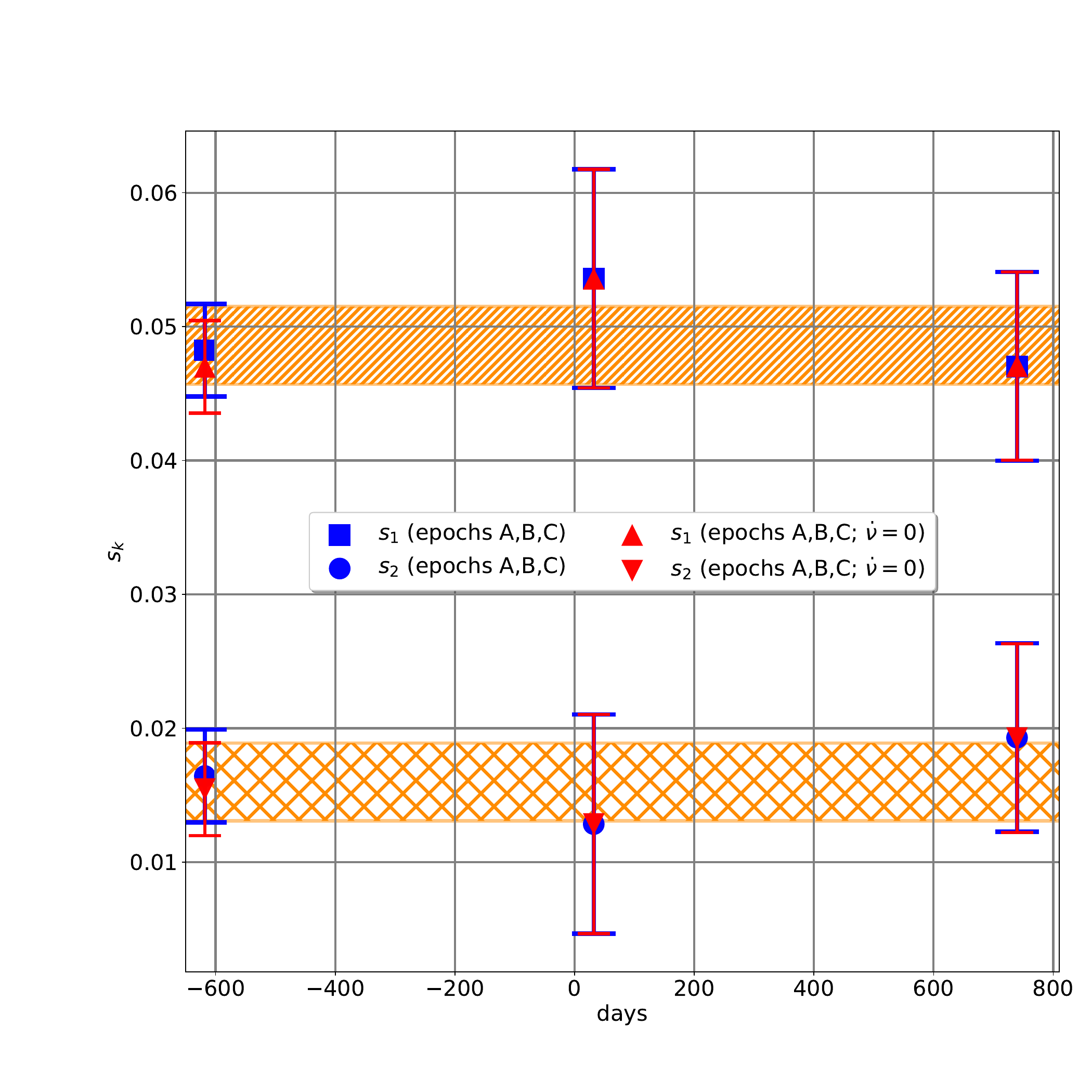}
\hspace{-0.3cm}
 \includegraphics[height=9.1cm]{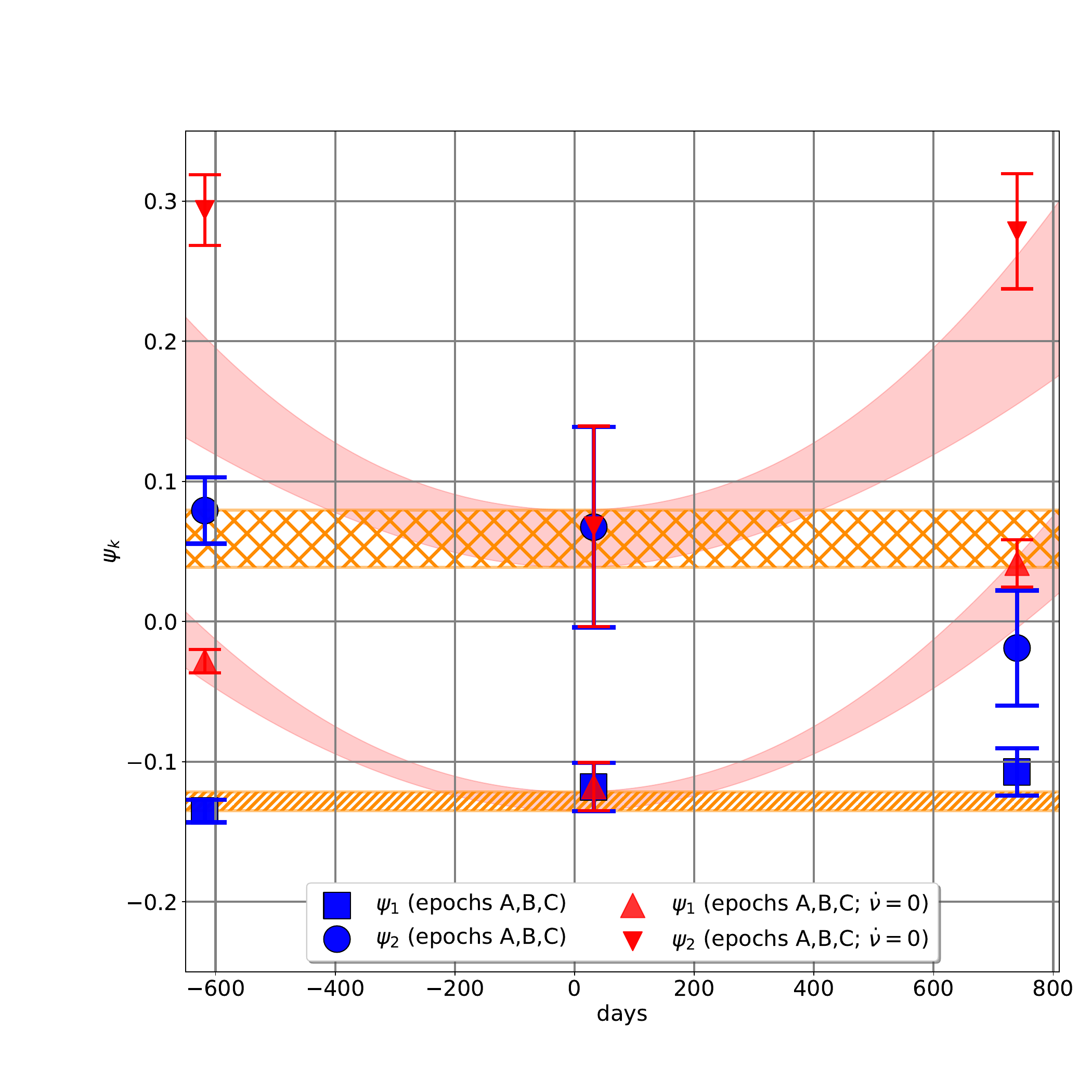}\\
\vspace{-0.2cm}
 \caption{The amplitudes and phases of two harmonics for the completely merged NICER data set (orange bands) in comparison to those of the merged epochs A, B, and C (blue square and circle symbols using the derived timing solution from the merged data $\nu=0.08794773254$\,Hz and $\dot{\nu}=-7.3 \times 10^{-17}$\,Hz/s). 
For comparison, we also show the expected result if $\dot{\nu}=0$ is assumed (red bands) and the measured values for this assumption, indicated with red triangle symbols. Error bars indicate $1\sigma$ uncertainties. 
\label{psi_s}}
\end{figure*}

The amplitudes and phases of the two relevant harmonics are plotted as functions of time for three merged epochs in Figure~\ref{psi_s}. We see that at the found timing solution the harmonic amplitudes $s_k$ do not change, and the harmonic phases $\psi_k$ remain about constant. This demonstrates that our observations are indeed phase-connected. 
In the same figure, we also show the $s_k$ and $\psi_k$ values obtained at the assumption that $\nudot = 0$.
While the harmonic amplitudes are virtually insensitive to the frequency derivative, the harmonic phases show significant relative shifts at the (wrong) $\nudot=0$, as expected. \\

The pulse profile of J0806's thermal X-ray emission is smooth and has a very low pulsed fraction.
For the energy range $0.23-1.0$\,keV, the observed max-to-min pulsed fraction is $p_{\rm amp, obs} =  (f_{\rm max} -f_{\rm min})/ (f_{\rm max}+f_{\rm min}) = 5.6\% \pm 0.3\%$, 
where the uncertainty was derived from Monte Carlo simulations.
Using the X-ray spectra of the merged data in the energy range $0.23-1.0$\,keV, the net source fraction is 90.8\%. Thus, the \emph{intrinsic} pulsed fraction of J0806 is $p_{\rm amp, int} =6.2\% \pm 0.3\%$.
For the other pulsed fraction definitions from Appendix C in \citet{Hare2021} 
we obtain $p_{\rm area, int} =6.8\% \pm 0.4\%$ and $p_{\rm rms, int} =4.0\% \pm 0.2\%$. 

\begin{deluxetable*}{lcccccc}
\tablecolumns{8}
\tablecaption{Estimates of $s_k$ and $\psi_k$ for our timing solution from Equation~\ref{timsol} (reference epoch MJD 59233.5) \label{skpsik}}
\tablehead{
\colhead{epochs} & \colhead{$s_1$} & \colhead{$\psi_1$} &  \colhead{$s_2$} & \colhead{$\psi_2$} & \colhead{$s_3$} & \colhead{$\psi_3$} \\
}
\startdata
mergeA & $0.0482 \pm 0.0035$ & $-0.1352 \pm 0.0081$ & $0.0164 \pm 0.0035$ & $ 0.079 \pm 0.024$ & $0.0031 \pm 0.0035$ & $0.24 \pm 0.13  $  \\
mergeB & $0.0536 \pm 0.0082$ & $-0.118 \pm 0.017  $ & $0.0128 \pm 0.0082$ & $ 0.067 \pm 0.072$ & $0.0129 \pm 0.0082$ & $0.154 \pm 0.071$  \\
mergeC & $0.0470 \pm 0.0070$ & $-0.107 \pm 0.017  $ & $0.0193 \pm 0.0070$ & $-0.019 \pm 0.041$ & $0.0101 \pm 0.0071$ & $0.239 \pm 0.078$  \\
\hline
A+B+C & $0.0486 \pm 0.0029$ & $-0.1283 \pm 0.0067 $  & $0.0160 \pm 0.0029$ & $0.059 \pm  0.020$ & $0.0053 \pm 0.0029$ & $0.216 \pm 0.061 $  \\
\enddata
\end{deluxetable*}

\section{Discussion} 
\label{sec:dis}
Our NICER observing campaign was planned as outlined in the Appendix~\ref{sec:planning}, starting  with a comparatively long observation and choosing the observing intervals of subsequent shorter observations in a way that ensured phase connection (see Appendix B).
Using this approach, we obtained our 7 unevenly spaced observation clusters spread over 3.9 years (see Table~\ref{NicerTab}).
Despite the large gaps between the observation epochs at the end of our campaign (e.g., 1.9 years between epochs 6 and 7), we were able to phase-connect all the observation clusters and confidently measure a rather small frequency derivative in a relatively short total exposure time of 182 ks.  
The good performance of this method with respect to the phase-connection is illustrated by Figure \ref{psi_s},
and it can also be seen in Table~\ref{Z2ks} where the difference between $Z^2_K$ of two subsequent rows
(lower half of table) is nearly\footnote{The values are not exactly the same due to differences in trial values and different reference times, and a lack of correction for timing noise from the non-pulsed background (see Equation~\ref{eq:Z-squared_sum_corrected}).} the value of $Z^2_K$ of that additionally included epoch.\\  
\begin{figure}[] 
\hspace{-0.9cm}
\includegraphics[height=8.7cm]{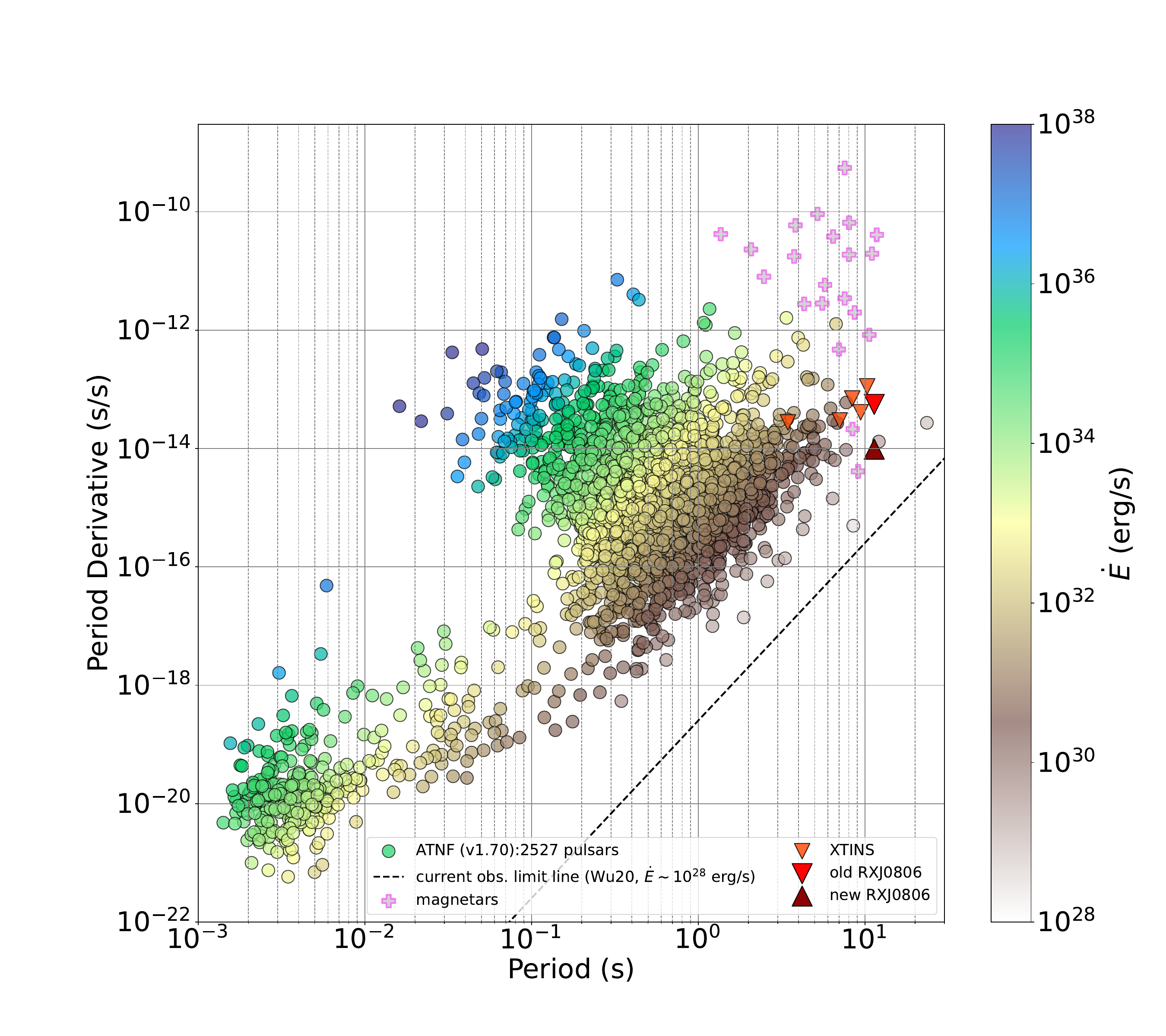}
 \caption{J0806 within the population of neutron stars. The \emph{left panel} shows the pulsars in the period--period-derivative diagram, color-coded by their spin-down energy. The old (uncertain) spin-down parameters by \citet{Kaplan2009} and our new values for J0806 are indicated by the downwards- and upwards-pointing red triangles, respectively. The current radio detection limits of steady pulsar emission according to \citet{Wu2020} are indicated by the dashed line.
\label{NSpopand0806}}
\end{figure}

Our NICER exposures from 2019 to 2023 allowed us to measure the rotation frequency of J0806 with a very high precision of $\sigma_\nu\sim 2\times 10^{-10}$\,Hz and evaluate its derivative, $\nudot = -(7.3\pm 1.2)\times 10^{-17}$ Hz\,s$^{-1}$, for the first time.   
The best-fit frequency derivative is a factor of 6 lower than the previous estimate, $\nudot= -4.3\pm 2.3 \times 10^{-16}$ Hz\,s$^{-1}$, by  \citet{Kaplan2009}. 
The pulse shape is satisfactorily described by two terms of the Fourier expansion (Table~\ref{skpsik}, Figure~\ref{pulseshape}).
This shape looks similar to the one described by \citet{Kaplan2009} and \citet{Haberl2004}.
This is also supported by the agreement of our background-corrected Fourier coefficients with those of \citet{Kaplan2009} within their $2\sigma$ uncertainties. In addition, we have checked that J0806's pulse shapes from the merged epochs A, B, C, and all the merged data agree with each other within their uncertainties, which is also seen from the nearly constant Fourier coefficients shown in Figure~\ref{psi_s}. Thus, the assumption of a stable pulse shape for our analysis is confirmed in retrospect. However, we note that variations in pulsed fractions, pulse shape and spectrum have been observed for another member of the XTINS, RX\,J0720.4-3125 \citep{Haberl2006}.\\

The two harmonics might be associated with a different temperature distribution in the two hemispheres. 
\citet{Vigano2014} have demonstrated that different asymmetric temperature distributions can be used to describe the spectral features of J0806's XMM-Newton data by simulating this neutron star's phase-averaged and phase-resolved spectra. 
However, although their models can reproduce the observed pulsed fraction, the model pulse profiles are very different from the observed one. This means that other models should be explored and compared with the observed phase-energy dependence, which requires a very large number of source counts. Our NICER timing data has about double the number of counts ($\sim 474$\,kcounts) in comparison to \citet{Kaplan2009} ($\sim 217$\,kcounts) for a smaller energy range. 
Hence, a phase-dependent spectral analysis could result in better constraints on geometry, temperature distribution, and absorption lines than it was possible before our NICER observations. 

Based on our timing solution, the phase 
uncertainty between the last XMM-Newton observation (MJD 54932.06) by \citet{Kaplan2009} 
and our reference epoch is $\sigma_{\Delta\phi}=[(\sigma_\nu^2 \Delta t^2 + \sigma_{\nudot}^2 \Delta t^4/4]^{1/2} =0.8$.
Thus, phase connection might be possible but is not guaranteed.
Count rates estimated for individual NICER observations are consistent with each other.
We also checked that the overall flux and spectral parameters estimated from the NICER observations are also consistent with 
those inferred from the XMM-Newton data. We defer a  detailed phase-resolved spectral analysis to future works.\\

The measured $\nudot$ value corresponds to the following pulsar parameters: spin-down power $\dot{E}=2.6\times 10^{29} I_{45}$\,erg\,s$^{-1}$ (where $I_{45}$ is the moment of inertia in unit of $10^{45}$\,g\,cm$^2$), magnetic field $B= 3.2\times 10^{19}  (P\dot{P})^{1/2} =1.1\times 10^{13}$\,G, and spin-down age $\tau = P/(2\dot{P})=18$ Myr (largest among XTINSs). 
As illustrated in Figure~\ref{NSpopand0806}, 
RX\,J0806 is among the pulsars with the lowest measured $\dot{E}$. The formal ``X-ray efficiency'',
$\eta_X \equiv L_X / \dot{E}=60$ (using the purely thermal X-ray luminosity of $1.6 \times 10^{31}$ erg s$^{-1}$, \citealt{Kaplan2009}) exceeds 1, similar to other XTINSs (for which $\eta_X$ varies from 0.2 to 68). 
The spin-down power is about a factor 6 lower than the 
previous estimate by \citet{Kaplan2009}. 
This new, lower value also has some implications for the parameters of a putative PWN, which is one of the possibilities to interpret the extended infrared emission discovered with the Hubble Space Telescope \citep{Posselt2018}. 
To balance the lowered $\dot{E}$, the magnetic field at the pulsar wind shock should be stronger by the factor of 6 in Equation (2) of \citet{Posselt2018} for the maximum electron energy.
Due to the large uncertainties, however, the PWN still remains a viable interpretation for the extended NIR emission.

\begin{acknowledgments}
We thank the anonymous referee whose remarks helped us to describe the used timing method more clearly.
This work was supported by NASA through the NICER mission
by NASA grants 80NSSC19K1447, 80NSSC21K0131, and 80NSSC22K1304. 
\end{acknowledgments}

%

\vspace{5mm}
\facilities{NICER}


\software{astropy \citep{2013A&A...558A..33A,2018AJ....156..123A},  
           HEAsoft \citep{HEAsoft2014}
          }


%

\appendix
\section{Measuring the UNCERTAINTIES OF FREQUENCY AND ITS DERIVATIVE
\label{sec:nu-nudot_uncertainties} }
\subsection{Basic formulae for the $Z_K^2$ statistic
\label{subsec:general_expressions}}
A timing observation of a source with a photon-counting detector provides a series of times of arrival, $t_i$, of the detected photons. If we know (or suspect) 
the source to be periodic, we can measure its frequency and frequency derivative (or search for pulsations) from a statistical analysis of phases of arrival,
\begin{equation}
    \phi_i =
    \phi_i(\nu,\nudot) =\phi_{\rm ref} + \nu (t_i-t_{\rm ref}) + \dot{\nu} (t_i-t_{\rm ref})^2/2\,,
\label{eq:event_phase}
\end{equation}
where $\nu$ and $\nudot$ are some assumed (trial) frequency and its derivative, and $\phi_{\rm ref}$ and $t_{\rm ref}$ are the reference phase and time, respectively.
Using the phases of arrival, one can calculate the $Z_K^2$ statistic (e.g., \citealt{Buccheri1983}),
\begin{equation}
    Z_K^2(\nu,\dot{\nu}) = \frac{N}{2} \sum_{k=1}^K s_k^2 
    = \frac{N}{2} \sum_{k=1}^K (a_k^2 + b_k^2)\,,
\label{eq:Zsquare_def}
\end{equation}
where 
$N$ is the total number of detected events (counts), $k$ is the Fourier harmonic number, 
$Ns_k^2/2$
is 
the power in the $k$th harmonic, $a_k$ and $b_k$ are the 
Fourier coefficients at the trial values of $\nu$ and $\nudot$:
\begin{equation}
    a_k =
    \frac{2}{N}\sum_{i=1}^N \cos2\pi k\phi_i =s_k \cos2\pi\psi_k
    , \quad\quad
    b_k  = 
    \frac{2}{N}\sum_{i=1}^N \sin2\pi k\phi_i
    = s_k \sin2\pi\psi_k\,,
\label{eq:Fourier_coeff}
\end{equation}
and $s_k$ and $\psi_k$ are the amplitude and phase of the $k$th harmonic (which can be considered as polar coordinates of the point with the Cartesian coordinates $a_k$, $b_k$). 
We emphasize that, being obtained from a set of observational data, all these quantities are prone to statistical (and perhaps systematic) errors. Besides, they depend on the assumed values of $\nu$ and $\nudot$ (or on $\nu$ only if $\nudot$ is not measurable in a given observation), which may be different from the actual values $\hat{\nu}$ and $\hat{\nudot}$ of the periodic signal\footnote{We mark all the ``actual'' properties of the parent signal (such as the count rate, harmonic amplitudes and phases, frequency) with a `hat' symbol.}.

Statistical uncertainties of 
$s_k$, $\psi_k$, and $Z_K^2$ can be found from the variances of the $a_k$ and $b_k$ 
coefficients (e.g., \citealt{Hare2021}),
\begin{equation}
\sigma^2_{a_k} = 
\frac{1}{N}\left(2+a_{2k} -  a_k^2\right), \qquad
\sigma^2_{b_k} = 
\frac{1}{N}\left(2-a_{2k} -   b_k^2\right)\,,
\label{eq:sigma_akbk}
\end{equation}
with the aid of error propagation. For instance,
\be
\sigma_{s_k}^2 = \frac{2}{N}\left[1 +\frac{1}{2} 
s_{2k} \cos\psi_{2k} \cos2\psi_k -\frac{1}{2}s_k^2 (\cos^4\psi_k +\sin^4\psi_k)\right]\,.
\label{eq:sigma_sk}
\ee
These equations show, in particular, that for small harmonic amplitudes the 
uncertainties of $a_k$, $b_k$ and $s_k$ only depends on the number of events, $\sigma_{a_k} \approx \sigma_{b_k} \approx \sigma_{s_k} \approx \sqrt{2/N}$.
Since the harmonic amplitudes usually decrease with increasing harmonic number, 
this approximation can be used to find the maximum number 
of harmonics to be used in the timing analysis -- e.g., 
$K_{\rm max}$ is the maximum number $K$ for which the condition $s_K > 3\sigma_{s_k} \approx 3\sqrt{2/N}$ is fulfilled.
In the case of RX\,J0806, all the merged NICER observations include $N=474,387$ events, which corresponds to $\sigma_{s_k} = 0.0029$. Comparing this uncertainty with the $s_k$ values (see Table \ref{skpsik}), we obtain $K_{\rm max}=2$.

The $Z_K^2(\nu,\nudot)$ statistic is commonly used to measure the frequency and its derivative, $\hat{\nu}$ and $\hat{\nudot}$, of a nearly periodic source.
For a given data set, the most likely frequency and frequency derivative estimates, $\nu_0$ and $\nudot_0$, 
correspond to the maximum of the $Z_K^2$ statistic on a $\nu$-$\nudot$ grid: $Z_K^2(\nu_0,\nudot_0) = Z_{K,\rm max}^2$.
If the frequency derivative is not measurable in this observation,  
then the most likely frequency is determined by the equation $Z_K^2(\nu_0)=Z_{K,\rm max}^2$.
The differences of the 
$\nu_0$ and $\nudot_0$
from the `true' values, $\hat{\nu}$ and $\hat{\nudot}$, are characterized by statistical uncertainties of the measured values, which can also be found with the aid of $Z_K^2(\nu,\nudot)$. We will consider some examples below.

\subsection{Estimating $\nu$ and $\nudot$ uncertainties for a sinusoidal signal}
\label{sec:sinusoidal}
\label{sec:A2}
Let us start from the simplest the case of a sinusoidal signal with a slowly changing frequency,
\be
\hat{f}(t) = \hat{\cal C} \left[1 + \hat{s} \cos 2\pi(\hat{\nu} t +\hat{\nudot} t^2/2 - \hat{\psi})\right]\,,
\label{eq:sin-signal}
\ee
where $\hat{\cal C}$ is the signal count rate averaged over pulsations, $\hat{s}$ 
and $\hat{\psi}$ are the signal amplitude and phase.
Statistical properties of such a signal for a given data set are fully described by the function $Z_1^2(\nu,\nudot)$ (also known as the Rayleigh test).

Statistical uncertainties of the frequency and its derivative 
of the sinusoidal signal, 
measured from a given data set,
are determined by shape and height of the $Z_1^2(\nu,\nudot)$ peak, which
in turn depend on shape and strength of the signal and on
observational setup (e.g., the number and durations of exposures included in the timing analysis). 
To explicitly connect the uncertainties with the height and shape of the peak, we can use the Taylor expansion of 
the function $Z_1^2(\nu,\nudot)$ near the peak's maximum:
\be
Z_1^2(\nu,\nudot) \approx Z_{1,\rm max}^2 [1 - A (\Delta\nu)^2 - B (\Delta\nudot)^2 - C (\Delta\nu) (\Delta\nudot)]\,,
\label{eq:ZKexpansion}
\ee
where
$\Delta\nu=\nu-\nu_0$, $\Delta\nudot = \nudot - \nudot_0$, $Z_{1,\rm max}^2 A = -(1/2) [\partial^2 Z_1^2(\nu,\nudot)/\partial\nu^2]$, $Z_{1,\rm max}^2  B= -(1/2) [\partial^2 Z_1^2(\nu,\nudot)/\partial\nudot^2]$, and $Z_{1,\rm max}^2  C= - [\partial^2 Z_1^2(\nu,\nudot)/(\partial\nudot \partial\nu)]$;
all the derivatives are taken at
$\Delta\nu=0$ and $\Delta\nudot=0$. 
The last term in Equation (\ref{eq:ZKexpansion}), related to correlation of $\nu$ and $\nudot$, can be eliminated by a proper choice of reference time, while the first two terms are related to the uncertainties of $\nu$ and $\nudot$.  
Cutting the peak by the horizontal plane at a height $Z_{1}^2 = Z_{1,\rm max}^2 - \Delta Z_1^2$ and projecting the cut onto the $\nu$-$\nudot$  plane, we obtain an elliptical confidence contour centered at $\nu_0$, $\nudot_0$, which is given by the equation:
\be
Z_{1,\rm max}^2 [A (\Delta\nu)^2 + B (\Delta\nudot)^2] = \Delta Z_1^2\,.
\label{eq:ellipse}
\ee
The semi-axes of this ellipse 
represent $\nu$ and $\nudot$ uncertainties at the confidence level determined by the $\Delta Z_1^2$ value:
\be
\delta\nu = \left(\frac{1}{A} \frac{\Delta Z_1^2}{Z_{\rm 1,max}^2}\right)^{1/2}, 
\qquad 
\delta{\nudot} = \left(\frac{1}{B} \frac{\Delta Z_1^2}{Z_{\rm 1,max}^2} \right)^{1/2}.
\label{eq:uncert_general}
\ee
The peak height $Z_{1,\rm max}^2$ can be found directly from the data, but in order  to find the uncertainties, we should know the coefficients $A$ and $B$, proportional to the second derivatives of $Z_1^2(\nu,\nudot)$ at the peak maximum, as well as the correspondence between the confidence level and the value of $\Delta Z_1^2$.

\subsubsection{Uninterrupted observation}
\label{sec:uninterrupted}
For a sufficiently long observation and a large number of detected photons, the function $Z_1^2(\nu,\nudot)$ can be written as\footnote{For the sake of clarity, here and throughout the Appendix, we do not consider the random noise contribution to $Z_K^2(\nu,\nudot)$. This contribution can be neglected for a sufficiently large signal power.}
\be
Z_1^2(\nu,\nudot) = \frac{N s^2}{2} |J(\nu,\nudot)|^2\,,
\label{eq:Z1_J}
\ee
where $N={\cal C}T_{\rm exp}$ is the number of detected counts, $T_{\rm exp}$ is the exposure time, ${\cal C}$ and
$s$ are the observed count rate and signal amplitude (equal to the observed pulsed fraction --- e.g., \citealt{Pavlov1999}), $Ns^2/2$ is the signal power, and $J(\nu,\nudot)$ is the modified Fourier transform of the signal $\hat{f}(t)$, normalized in such a way that $|J(\nu_0,\nudot_0)|^2 = 1$.

For an uninterrupted observation of duration $T$, we have $T_{\rm exp} = T$ and
\be
J(\nu,\nudot) = \frac{1}{T}\int_{-T/2}^{T/2} \exp\left[-2\pi i (\Delta\nu t +\Delta\nudot \frac{t^2}{2} + \psi)\right]\, dt\,.
\label{eq:J-general}
\ee
For $\nudot= \nudot_0$, we obtain a 1D profile of $Z_1^2$ along the frequency axis,
\be
Z_1^2(\nu,\nudot_0) = \frac{Ns^2}{2} {\rm sinc}^2(\pi\Delta\nu\,T)\,,
\label{eq:sinc}
\ee
where ${\rm sinc}(x) = (\sin x)/x$ -- a well-known formula for the Fourier power of a purely sinusoidal signal (e.g., \citealt{Leahy1983}).
At $\nu=\nu_0$, the 1D profile along the $\nudot$ axis is 
\be
Z_1^2(\nu_0,\nudot) = \frac{Ns^2}{2y^2} \left[C^2(y) + S^2(y)\right]\,,
\label{eq:fresnel}
\ee
where $y=(\pi |\Delta\nudot|)^{1/2} T/2$ and $C(y)=\int_0^y \cos(u^2)\,du$, $S(y)=\int_0^y \sin(u^2))\,du$ are the Fresnel integrals. The 2D $Z_1^2(\nu,\nudot)$ peak can also be expressed in terms of Fresnel integrals of a more complicated arguments (Pavlov et al., in prep.).

Using Taylor expansions over powers of $\Delta\nu$ and $\Delta\nudot$ of the right-hand sides of Equations (\ref{eq:sinc}) and (\ref{eq:fresnel}), respectively,
we obtain the coefficients in Equations (\ref{eq:uncert_general}),  $A=\pi^2 T^2/3$ and $B=\pi^2T^4/180$, and the uncertainties
\be
\delta\nu = \frac{1}{\pi T} 
\sqrt{\frac{3 \Delta Z_1^2}{Z_{\rm 1,max}^2}} = \frac{1}{\pi T s}\sqrt{\frac{6 \Delta Z_1^2}{N}}\simeq
0.78\,{\cal C}^{-1/2} s^{-1} T^{-3/2} \left(\Delta Z_1^2\right)^{1/2}\,,
\label{eq:deltanu_gen}
\ee
\be
\delta\nudot = \frac{6}{\pi T^2}\sqrt{\frac{5 \Delta Z_1^2}{Z_{\rm 1,max}^2}} =\frac{6}{\pi T^2 s}\sqrt{\frac{10 \Delta Z_1^2}{N}} \simeq 6.0\, {\cal C}^{-1/2} s^{-1} T^{-5/2} \left(\Delta Z_1^2\right)^{1/2}\,.
\label{eq:deltanudot_gen}
\ee
If we assume $\Delta Z_1^2 =1$, then (\ref{eq:deltanu_gen}) and (\ref{eq:deltanudot_gen}) turn into the
familiar equations for standard deviations 
(\citealt{Hare2021,Chang2012,Ransom2002}):
\be
\sigma_\nu = (\sqrt{3}/\pi) T^{-1} (Z_{1,\rm max}^2)^{-1/2}  
\simeq 0.78\, {\cal C}^{-1/2} s^{-1} T^{-3/2}\,,
\label{eq:nu_error}
\ee
\be
\sigma_{\nudot} = (6\sqrt{5}/\pi) T^{-2} (Z_{1,\rm max})^{-1/2} 
\simeq 6.0\, {\cal C}^{-1/2} s^{-1} T^{-5/2}\,.
\label{eq:nudot_error}
\ee
These estimates 
have been confirmed by Monte-Carlo simulations \citep{Chang2012,Hare2021}, which supports the 
conjecture that the 68\% confidence level for the uncertainties of frequency and its derivative corresponds to $\Delta Z_1^2=1$.

\subsubsection{The case of several observations and the empirical approach for estimating $\nu$ and $\nudot$ uncertainties.}
\label{sec:mult_obs}
If several observations of a source with constant timing properties
are phase-connected, they 
can be treated as a single observation with gaps, and the function $Z_1^2(\nu,\nudot)$ can be calculated in a way similar to that used for an uninterrupted observation. For instance, for a sinusoidal signal,
one can modify Equations (\ref{eq:Z1_J}) and (\ref{eq:J-general})  
by replacing the integral over time by several integrals over actual exposure intervals and changing the $1/T$ factor by $1/T_{\rm exp}$.
The presence of gaps leads to appearance of multiple peaks of $Z_1^2(\nu,\nudot)$; the main peak (usually the highest one) corresponds to the actual (most probable) timing parameters, while the other peaks correspond to aliases. Different aliases correspond to different integer phase shifts with respect to the main peak during time intervals $\Delta t$ in the dataset, with 
$\Delta t$ representing various differences and sums of gap durations and exposure times.  
Thus, the phase shift of an alias is 
$\Delta\phi = \Delta\nu\,\Delta t + \Delta\nudot\, (\Delta t)^2/2 \simeq n$, where $\Delta\nu$ and $\Delta\nudot$ are the coordinates of the alias on the $\nu$-$\nudot$ map, and $n$ ($\neq 0$) is an integer.\\

For instance, for the simplest case of two observations of the same duration $T_{\rm exp}/2$, separated by a large gap $T_{\rm gap}=T-T_{\rm exp}$, we obtain (for $T_{\rm exp}\ll T$)
\be
Z_1^2(\nu,\nudot_0) = ({\cal C}T_{\rm exp} s^2/2)\, {\rm sinc}^2(\pi\Delta\nu T_{\rm exp}/2)\, \cos^2(\pi \Delta\nu T)\,.
\label{eq:one_gap_Zsquare}
\ee
The factor $\cos^2(\pi \Delta\nu T)$ is responsible for multiple narrow peaks of $Z_1^2(\nu)$, with short periodic spacings 
of $T^{-1}$ along the frequency axis.
This structure is modulated by the function ${\rm sinc}^2(\pi\Delta\nu T_{\rm exp}/2)$ with a much larger periodic spacing 
 of $\sim 2/T_{\rm exp}$.
The highest peak, 
with height $Z_{\rm 1,max}^2={\cal C}T_{\rm exp} s^2/2$, occurs at $\nu=\nu_0$, and its half-width at the level of $Z_{\rm 1,max}^2 - 1$ 
is 
\be
\sigma_\nu = (\pi T)^{-1} \left(Z_{1,\rm max}^2\right)^{-1/2} = 
0.45\, {\cal C}^{-1/2} s^{-1} T^{-1} T_{\rm exp}^{-1/2}\,.
\label{eq:sigmanu_gap}
\ee
 The other narrow peaks with similar widths are the alias peaks caused by the gap. Although the alias peaks are lower than the main one, the differences between the heights of the main peak and the nearest alias peaks become too small when $T/T_{\rm exp}$ becomes so large that the phase connection between the two observation is lost (see Appendix \ref{sec:planning}). Therefore, although
$\sigma_\nu\propto T^{-1}$, 
the frequency cannot be measured with a very high precision by choosing a very long gap between the observations. 

Comparing Equations (\ref{eq:sigmanu_gap}) 
and (\ref{eq:nu_error}),
we see that, at least for a sinusoidal signal, the dependencies of $\sigma_\nu$ 
on the signal parameters, exposure time and total time span are qualitatively the same for both an uninterrupted observation and an observation with a gap (taking into account that $T_{\rm exp}=T$ for an uninterrupted observation). It suggests that in a general case of several observations with different exposures and time gaps we should expect a universal dependence 
\be
\sigma_\nu \propto ({\cal C} T_{\rm exp})^{-1/2} s^{-1} T^{-1}\,  
\ee
with the numerical proportionality coefficient depending on observational setup. This relationship is useful for crude estimates of the frequency 
uncertainty 
and for planning of timing observations (see Appendix \ref{sec:planning}).

In the case of many individual observations of different durations the multi-peak $Z_1^2(\nu,\nudot)$ dependence becomes very complicated (see Figure 1), and the use of very cumbersome analytical expressions for estimating the $\nu$ and $\nudot$ uncertainties becomes impractical. However, the above-considered examples suggest that one can use a much simpler 
approach for this purpose. Once, for a given data set, the ``correct'' peak of $Z_1^2(\nu,\nudot)$ is identified,
one can estimate the $1\sigma$ uncertainties from the equations
\be
Z_1^2(\nu_0\pm \sigma_\nu, \nudot_0) = Z_{1,\rm max}^2 -1,\qquad Z_1^2(\nu_0, \nudot_0\pm \sigma_{\nudot})
=Z_{1,\rm max}^2 -1.
\label{eq:empirical_uncertainties}
\ee
We have used this {\em empirical approach} in timing analysis of XMM-Newton observations of pulsar B1055--52 \citep{Vahdat2024}.
The estimates of the $\nu$, $\nudot$ uncertainties 
inferred 
from the NICER data RX\,J0806 are also based on this 
approach, with some modification to account for the presence of the frequency harmonic.

\subsection{Effect of signal harmonics on the frequency uncertainty}
Let us now consider the case when the signal contains harmonics $k>1$:
\be
\hat{f} = \hat{\cal C} \left[1 +\sum_{k=1}^{K} \hat{s}_{k} \cos 2\pi(k\hat{\nu} t + \hat{\nudot} t^2/2 - \hat{\psi}_{k})\right]\,.
\ee
In this case the $Z_1^2(\nu)$ statistic\footnote{Since the presence of harmonics does not affect $\nudot$ measurement, we  consider only the frequency dependence here.}
 has peaks not only near the fundamental frequency
 $\hat{\nu}$ but also at its harmonics $k\hat{\nu}$.
The peaks of the $Z_1^2(\nu)$ near $k\hat{\nu}$ 
have the heights $Ns_k^2/2$, while their shapes are the same as the shape of the peak at the fundamental frequency.
Similar to the case of 
sinusoidal signal, the uncertainty of the frequency $\nu_0^{(k)}
\simeq k\hat{\nu}$, measured from the $k$th peak, is 
given by Equation (\ref{eq:nu_error}), which means that the corresponding uncertainty of $\hat{\nu}$ is a factor of $k$ smaller:   
\be
\sigma_\nu^{(k)}= \frac{1}{\pi T ks_k} \sqrt{\frac{6}{N}}\,.
\ee
Now we can say that we have $K$ measurements 
of $\hat{\nu}$, with different uncertainties given by the above equation. Then, using the usual rule for calculating the variance of a mean value obtained from several observations,
we obtain the resulting $1\sigma$ uncertainty:
\be
\sigma_{\nu} = \frac{\sqrt{6}}{\pi T} \left(N\sum_{k=1}^K k^2 s_k^2\right)^{-1/2}
= 0.78\, {\cal C}^{-1/2} \left(\sum_{k=1}^K k^2 s_k^2\right)^{-1/2} T^{-3/2}\,.
\label{eq:freq-uncert-harmonics-nogap}
\ee 
Equation (\ref{eq:freq-uncert-harmonics-nogap}) turns into Equation 
(\ref{eq:nu_error}) if contributions of the $k>1$ harmonics are negligible.
Taking those harmonics into account
reduces the frequency uncertainty by a factor of 
\be
g =\left(1+\sum_{k=2}^K k^2 (s_k/s_1)^2\right)^{1/2}.
\label{eq:g-factor}
\ee
For instance, $g\approx 1.2$ for the 0.23--1.0 keV pulsations of RX\,J0806 in our NICER observations. Thus, taking higher harmonics into account allows one to measure the frequency more precisely, albeit the improvement may be small.
Moreover, once the $\sigma_\nu$ is measured from $Z_1^2(\nu)$, it can be easily corrected for the presence of higher harmonics by dividing over the $g$-factor, given by Equation (\ref{eq:g-factor}).

\subsection{Uncertainties at different confidence levels} 
We have derived a few equations for $1 \sigma$ uncertainties, $\sigma_\nu$ and $\sigma_{\nudot}$, of frequency and its derivative. It would also be useful to know the uncertainties at different confidence levels. In principle, they could be obtained from Equations (\ref{eq:uncert_general}), but the correspondence between the $\Delta Z_1^2$ value and the confidence level is not immediately clear\footnote{The statement in \citet{Buccheri1983} that ``The variable $Z_n^2$ has a probability density distribution equal to that of a $\chi^2$ with $2n$ degrees of freedom'' is only applicable to timing noise.}. 
However, we can infer this correspondence with the aid of the Bayesian posterior probability density derived from the likelihood marginalized over `nuisance parameters' \citep{Bretthorst1988}.
It follows from Chapter 2 of that book that for a purely sinusoidal signal (Equation \ref{eq:sin-signal} with $\nudot_0=0$), the normalized probability density in the vicinity of $\nu=\nu_0$ 
obeys a Gaussian distribution
\be
p(\nu) = \frac{1}{\sqrt{2\pi} \sigma_\nu} \exp\left[-\frac{(\nu-\nu_0)^2}{2\sigma_\nu^2}\right]\,,
\label{eq:gaussian}
\ee
where $\sigma_\nu$ is given by Equation (\ref{eq:uncert_general}) for $\delta\nu$ with $\Delta Z_1^2 =1$. 
It means that, as long as Equation (\ref{eq:gaussian}) is applicable, an $a \sigma$ confidence level corresponds to $\Delta Z_1^2 = a^2$. In other words, the 
confidence levels for the frequency uncertainty of, e.g., 68.3, 90.0, 99.0, 99.73 percent correspond to $Z_1^2 \approx 1$, 2.7, 6.6, and 9, respectively.

If $Z_1^2$ depends on both $\nu$ and $\nudot$, then the natural generalization of Equation (\ref{eq:gaussian}) is
\begin{equation}
p(\nu,\nudot) = \frac{1}{2\pi \sigma_\nu \sigma_{\nudot}} \exp\left[-\frac{(\nu -\nu_0)^2}{2 \sigma_\nu^2} -\frac{(\nudot -\nudot_0)^2}{2 \sigma_{\nudot}^2} \right]\,,
\label{eq:2D-gaussian}
\end{equation}
assuming the reference epoch is such that $\nu$ and $\nudot$ are not correlated.
We can use this equation to calculate confidence levels corresponding to different contours 
$Z_1^2(\nu,\nudot) = Z_{\rm 1,max}^2 - \Delta Z_1^2  $ in the $\nu$-$\nudot$ plane.  For a given $\Delta Z_1^2$, the contour is given by the equation
\be
\frac{(\nu -\nu_0)^2}{\sigma_\nu^2} +\frac{(\nudot - \nudot_0)^2}{\sigma_{\nudot}^2} = \Delta Z_1^2\,,
\ee
which is the ellipse with semi-axes $\sigma_\nu \sqrt{\Delta Z_1^2}$ and $\sigma_{\nudot}\sqrt{\Delta Z_1^2}$. If we introduce new variables $x=\Delta\nu/\sigma_\nu$ and $y=\Delta{\nudot}/\sigma_{\nudot}$, then the contour is a circle with a radius $\sqrt{\Delta Z_1^2}$:
\be
x^2 + y^2 = \Delta Z_1^2\,.
\ee
The probability that $x$ and $y$ are within this circle is
\be
P (x^2+y^2 \leq Z_1^2)  =
 1 - \exp(-\Delta Z_1^2/2)\,.
\ee
This allows us to calculate $\Delta Z_1^2$ corresponding to a given probability
\be
\Delta Z_1^2 = 2\ln\frac{1}{1-P}\,.
\ee
For instance, $\Delta Z_1^2 = 2.30$, 4.61, 9.21 and 11.62  for $P=0.68$, 0.90, 0.99 and 0.997, respectively. Note that these 
are confidence contours for 2 parameters of interest; they correspond to larger $Z_1^2$ than the 1D Gaussian values. 

\subsection{Summary of the empirical approach}
To summarize, the empirical approach to estimating the $\nu$ and $\nudot$ uncertainties includes the following steps. 
\begin{itemize}
\item 
For a given set of phase-connected observations, calculate the statistic $Z_1^2(\nu,\nudot)$ (or just $Z_1^2(\nu)$ if $\nudot$ is not measurable) on an appropriate  grid of $\nu$-$\nudot$ (or just $\nu$), choosing the most convenient reference epoch near the center of the total time span.
\item
Find the highest peak of $Z_1^2(\nu,\nudot$) and 
determine the most likely frequency $\nu_0$ and frequency derivative $\nudot_0$, corresponding to the peak's maximum.
\item 
Estimate the $\nu$ and $\nudot$ uncertainties at the confidence level of 68\% (for one parameter of interest) from Equation (\ref{eq:empirical_uncertainties})
with $\Delta Z_1^2 =1$, 
in the one-harmonic approximation.
\item 
Calculate the Fourier coefficients $s_k$ and $\psi_k$, and their $1\sigma$ uncertainties at $\nu = \nu_0$, $\nudot=\nudot_0$ for several consecutive harmonics $k\geq 1$. Determine the maximum number $K$ of statistically significant harmonics for which $s_K > \cal{A} \sigma_{s_k}$, where the value of ${\cal A}$ depends on the desired confidence level (e.g., ${\cal A}=3$ for the confidence level of 99.7\%). Use $s_K$ and $\psi_K$ to plot the pulse profile (folded light curve).
\item 
Apply the higher-harmonics correction (Equation \ref{eq:g-factor}) to the frequency uncertainty. 
\end{itemize}

\section{Phase-connecting observations by maximizing the $Z$-squared statistics
\label{sec:Zn2app}}
If we have $M$ 
observations of the same source with large time gaps between 
them, we can measure the frequencies 
$\nu_m$ ($m=1, 2, \ldots , M$) 
at which the $Z$-squared statistics for separate observations, 
$Z_{K,m}^2=(N_m/2)\sum_{k=1}^K (a_{k,m}^2 + b_{k,m}^2)$, 
are maximal, fit the sequence of $\nu_m$ with a linear function of time, and determine $\nudot$ from the slope of this line. 
This approach provides an
incoherent measurement of $\nudot$ (the pulsation phases $\psi_{k,m}$ in separate observations are not connected).

An alternative way to measure 
the frequency derivative is maximizing $Z_K^2(\nu,\nudot)$
for a combined data set, i.e., simultaneously for all the $M$ observations. 
The Fourier coefficients for the combined data set can be written as
\be
a_k = \frac{2}{N}\sum_{m=1}^M \sum_{i_m=1}^{N_m} \cos 2\pi k \phi_{i_m} = \sum_{m=1}^M \frac{N_m}{N} a_{k,m} = \sum_{m=1}^M \frac{N_m}{N} s_{k,m} \cos 2\pi \psi_{k,m}
\label{eq:a_k_combined}
\ee
\be
b_k = \frac{2}{N}\sum_{m=1}^M \sum_{i_m=1}^{N_m} \sin 2\pi k \phi_{i_m} = \sum_{m=1}^M \frac{N_m}{N} b_{k,m} = \sum_{m=1}^M \frac{N_m}{N} s_{k,m} \sin 2\pi \psi_{k,m}
\label{eq:b_k_combined}
\ee
where $N=\sum_{m=1}^M N_m$. Substituting these expressions into Equation (\ref{eq:Zsquare_def}), we obtain
\begin{eqnarray}
   Z_K^2(\nu,\nudot)&= &\frac{1}{2N}\sum_{k=1}^K \sum_{m,m'=1}^M N_m N_{m'} s_{k,m}s_{k,m'} \cos 2\pi (\psi_{k,m}-\psi_{k,m'})\\
   & = & \frac{1}{2N}\sum_{k=1}^{K} \left[\left(\sum_{m=1}^M N_m s_{k,m}\right)^2 - 4\sum_{m=1}^M \sum_{m'=1}^{m-1} (N_m s_{k,m})(N_{m'}s_{k,m'})\sin^2\pi(\psi_{k,m}-\psi_{k,m'})\right]
   \,.
\label{eq:Z-square_combined}
\end{eqnarray}

The quantities $s_{k,m}$, $s_{k,m'}$, $\psi_{k,m}$ and $\psi_{k,m'}$ depend on the trial frequency and its derivative. Obviously, maximizing $Z_K^2(\nu,\nudot)$ implies minimizing the phase differences $\psi_{k,m}-\psi_{k,m'}$, i.e., the phase connection of the separate observations.
If the `actual' amplitudes $\hat{s}_k$ and phases $\hat{\psi}_k$ of harmonics remain the same (i.e., no glitches or substantial flux changes) within the time span of the $M$ observations, then, at the actual frequency and its derivative, we expect $s_{k,m'}= s_{k,m} = s_k$, $\psi_{k,m'} = \psi_{k,m'} = \psi_k$, and
\be
Z_{K;\rm max}^2 = \sum_{m=1}^M Z_{K,m;\rm max}^2\,. 
\label{eq:sum_observations}
\ee
It means that 
at the ideal phase connection 
the maximum of $Z_K^2$ for the entire data set is equal to the sum of $Z_K^2$ maxima found for the data subsets.  
It, however, can be somewhat lower in a real situation because of statistical and/or systematic errors.
In addition, 
in the derivation of Equation (\ref{eq:sum_observations}) we neglected the contribution of timing noise from the non-pulsed background to the $Z_{K}^2$ values. The expected (mean) value of this contribution to each of the $Z_K^2$  is equal to $2K$, which means that for timing noise-corrected $Z_K^2$ Equation (\ref{eq:sum_observations}) turns into 
\be
Z_{K;\rm max}^2 = \sum_{m=1}^M Z_{K,m;\rm max}^2 - 2K(M-1)\,.
\label{eq:Z-squared_sum_corrected}
\ee
This correction explains most of differences in Table \ref{Z2ks} between the {\em uncorrected} $Z_{K,\rm max}^2$ values for a data set and the sum of $Z_{K, \rm max}^2$ values for separate subsets.

\section{Planning a series of phase-connected observations of a pulsar
\label{sec:planning}}
To measure a frequency derivative 
in a photon-counting observation (or a series of observations) of a pulsar, not only one should detect enough pulsar counts, but also the time span of the observation(s) should be sufficiently long.  
The required time span is usually too long for an uninterrupted observation, and a number of much shorter observations have to be carried out. To plan such a program, one needs to optimize the durations of the separate observations and the time gaps between them.
An important criterion for this choice is the requirement of phase connection between separate observations, which greatly increases the accuracy of timing analysis. If the condition of phase connection is fulfilled, we can exactly count the number of pulsation cycles within the entire time span and determine phases of all events (counts) detected over the time span. 

One can consider two observations as phase-connected when the phase difference 
between events from the two observations is known with a precision better than a fraction $\alpha$ of the pulsation cycle.
To ensure phase connection for a series of observations, the time $t_{m+1}$ between the observation epoch 
$m+1$ and the preceding epoch $m$
should be such that the phase uncertainty $\delta\phi_{m+1}$, reached by the epoch $m+1$, is smaller than $\alpha$. As long as the $m$ observations remain insensitive to $\nudot$, or $t_{m+1}$ does not exceed the  time span $T_{{\rm span},m}$ from the start of epoch 1 to the end of epoch $m$, the main contribution to the phase uncertainty comes from the uncertainty in the frequency: 
\be
\delta\phi_{m+1} = t_{m+1} \delta\nu_{m} < \alpha\,, \qquad 
\delta\nu_m = a T_{\rm span,m}^{-1} \left(N_m\sum_{k=1}^K k^2 s_k^2\right)^{-1/2}\,,
\label{eq:delta_phi}
\ee
where 
$\delta\nu_m$ is the frequency uncertainty, 
estimated from the entire set of preceding (phase-connected) observations.
The frequency uncertainty is inverse proportional to the total time span $T_{{\rm span},m}$ from the start of epoch 1 to the end of epoch $m$, 
and to the square root of the total number of counts $N_m = {\cal C} \sum_{m'=1}^m T_{{\rm exp}, m'}$ collected in the entire data set (${\cal C}$ is the source count rate).
Since the contribution of higher harmonics is small, at least in our case, the sum over harmonics can be approximated as $\sum_{k=1}^K k^2 s_k^2 \approx s_1^2 \approx p_{\rm obs}^2$, where $p_{\rm obs}$ is the observed pulsed fraction.
The coefficient $a\sim 1$ depends on 
durations of separate observations and time gaps (see Section \ref{sec:mult_obs}).
For planning purposes, we 
chose 
$a=1$ and a conservative $\alpha = 0.2$. The values of the pulsation amplitude ($\approx$ pulsed fraction), $s_1=0.05$, and the NICER count rate $C=2.2$ counts s$^{-1}$ in the  energy range 0.23--1.0 keV, optimal for timing, were chosen from previous XMM-Newton observations of our target.
To ensure that phase connection is not lost in the beginning of such a multi-observation program, 
we chose the first observation (A1 in Table 1) considerably longer than the others. 

\bibliography{rxj0806N}{}
\bibliographystyle{aasjournal}



\end{document}